\documentclass[twocolumn,prd,groupedaddress,10pt]{revtex4}
\usepackage{graphicx}
\usepackage{amsmath}
\usepackage{dcolumn}
\usepackage{bm}
\usepackage{float}

\usepackage{tikz}
\usetikzlibrary{trees}
\usetikzlibrary{decorations.pathmorphing}
\usetikzlibrary{decorations.markings}

\def\lsim{\mathrel{\mathstrut\smash{\ooalign{\raise2.5pt\hbox{$<$}\cr\lower2.5pt\hbox{$\sim$}}}}}
\def\gsim{\mathrel{\mathstrut\smash{\ooalign{\raise2.5pt\hbox{$>$}\cr\lower2.5pt\hbox{$\sim$}}}}}

\def\baa{\begin{align}}
\def\eaa{\end{align}}
\def\be{\begin{equation}}
\def\ee{\end{equation}}
\def\bea{\begin{eqnarray}}
\def\eea{\end{eqnarray}}

\begin{document}

\title{Static structure of chameleon dark matter as an explanation of dwarf spheroidal galaxy cores}

\author{Prolay Krishna Chanda, Subinoy Das }
\email{prolay.krishna@iiap.res.in, subinoy@iiap.res.in}
\affiliation{Indian Institute of Astrophysics, Bangalore, 560034, India}


\begin{abstract} 

We propose a novel mechanism that explains the cored dark matter density profile in recently observed dark matter rich dwarf spheroidal galaxies. In our scenario, dark matter particle mass decreases gradually as a function of distance  towards the center  of a dwarf  galaxy due to its interaction with a chameleon scalar. At closer distance towards the Galactic center  the strength of attractive scalar fifth force becomes much stronger than gravity and is balanced by the Fermi pressure of the dark matter cloud; thus, an equilibrium static configuration of the dark matter halo is obtained.  Like the case of soliton star or fermion Q-star, the stability of the dark matter halo is obtained as the scalar achieves a static profile and reaches an asymptotic value away from the Galactic center. For simple scalar-dark matter interaction and quadratic scalar self-interaction potential, we show that dark matter behaves exactly like cold dark matter (CDM) beyond a few $\rm{kpc}$ away from the Galactic center but at closer distance it becomes lighter and Fermi pressure cannot be ignored anymore. Using Thomas-Fermi approximation, we numerically solve the radial static profile of the scalar field, fermion mass and  dark matter energy density as a function of distance. We find that for fifth force mediated by an ultralight scalar, it is possible to obtain a flattened dark matter density profile towards the Galactic center. In our scenario, the fifth force can be neglected at distance  $ r \geq 1\, \rm{kpc}$  from the Galactic center and dark matter can be simply treated as heavy nonrelativistic particles  beyond this distance, thus reproducing the success of CDM at large scales. 
 \end{abstract}

\maketitle

\section{ Introduction:}

   In spite of extensive research the nature of dark matter (DM)still remains a mystery. Though its existence is confirmed only through its gravitational effect, it is widely accepted that modified gravity cannot be a substitute for particle dark matter, especially when one would like to reproduce the results from large scale cosmological observations like cosmic microwave background (CMB), baryon acoustic oscillations (BAO),  and many more. For the search for particle dark matter in direct, indirect and collider detection, weakly interacting supersymmetric cold dark matter (CDM) has remained in the forefront as the most popular candidate. But its nondetection in spite of extensive research as well as  conflicting results between different direct and indirect detection experiments \cite{Hooper:2013cwa}  could be a strong hint to look for candidates of dark matter beyond the CDM paradigm. Other strong motivations to look beyond CDM originate from the long-standing small scale issues when one tries to match CDM  N-body simulations predictions with the galactic observations \cite{Weinberg:2013aya}. Though in large scale observations (like CMB and BAO) CDM is amazingly successful in matching the observed data and  predictions from linear perturbations, in the nonlinear regime small scale issues like the ``satellite problem" \cite{Klypin:1999uc, Moore:1999nt}, ``core-cusp" problem \cite{deblok(2010), Walker(2011)}, and 
  ``too big to fail problem" \cite{boylan(2011), garrison(2014)}  remain a strong challenge to the CDM paradigm. Though the recently incorporated baryonic feedback process has been proposed as a possible solution  to these challenges, the situation remains unclear and some recent works suggest that the above problems still may persist even if baryonic feedback is taken into account \cite{Onorbe:2015ija, Sawala:2014xka, Pawlowski:2015qta}. But more prominently, recent observations of the cored density profile of the dark matter halo in small low-surface-brightness (LSB) galaxies (for example F568-3) or satellite dSph galaxies like Fornax and Sculptor exaggerate the CDM core-cusp problem as these dSph galaxies are dark matter rich  with very high mass to light ratio. So one would not expect baryonic feedback to be that effective for N-body simulation  in explaining the cored profile in these objects. \par As a solution to the above issues of CDM, an alternative dark matter candidate made up of $\rm{keV}$ sterile neutrino was proposed. Because of its light mass compared to $\rm{GeV}$ CDM, this warm dark matter (WDM) particles  free-stream  in the early epoch of structure formation, thus suppressing matter power in small scales. Because of this lack of power in the small scale, when put into N-body simulation, WDM has shown some promise to solve the CDM challenges. But it is also not free of problems---  in fact WDM does too much of a good job in erasing satellite galaxies in N-body simulation and success of WDM in explaining the CDM challenges is limited to a narrow (fine-tuned) band of thermal WDM mass (between 1.5 \rm{keV} and 2 \rm{keV})\cite{lovell(2012)}. On top, this range of mass is claimed to be ruled out if one takes Lyman-alpha data seriously \cite{ Baur:2015jsy}. Also it is instructive to note that the  core-cusp issue is not fully resolved by WDM simulation \cite{Macci(2012)}.\\
 \,\,\,\,Recently another popular alternative has drawn lot of interest where dark matter is formed after big bang nucleosynthesis but considerably before matter radiation equality. This late forming dark matter can appear from Ultra light axion (ULA) \cite{Marsh:2015xka,Hui:2016ltb,Blas:2016ddr, Banik:2017ygz, Navarrete:2017txh } in the string axiverse \cite{Arvanitaki:2009fg} scenario or also from the extended neutrino sector \cite{Sarkar:2014bca,Das:2006ht, Berlin:2016woy}. High resolution N-body simulation has been performed on both the models and the results \cite{Schive:2014dra,Agarwal:2014qca} seem to solve the core cusp and some of the other issues of CDM. For ultralight axion DM, on top of the success of N-body simulation, an analytical solution incorporating quantum effects \cite{Marsh:2015wka} at small distances below de Broglie wave length of the ULA seems to match the  N-body cored density profile. But this scenario is also not free from observational challenges as the recently measured abundance of ultrafaint lensed galaxies at $z \simeq 6$ in the Hubble Frontier Fields might provide stringent constraints on the success of ULA dark matter in explaining dSph cores\cite{Menci:2017nsr}. Another dark matter candidate, which was proposed very recently \cite{Berezhiani:2015bqa}, claims to solve the core-cusp issue at small scales due to the superfluidic effect of the scalar-dark matter condensate at the small scale while replicating the success of CDM at the large scale.
 \par  Here in this work, we propose a physical mechanism for the first time that even a CDM candidate can predict the cored dark matter profile when it has an interaction with a long range scalar. CDM interacting with the ultralight cosmological scalar is common in much recent work \cite{Friedman:1991dj,Boyle:2001du, Kumar:2013ecq}, which may have its origin in string theory. Though originally introduced as a large scale modification of gravity to explain dark energy, recently, chameleon scenario has drawn lot of interests in explaining  small scale phenomena \cite{Brax:2017wcj, Sakstein:2016ggl, Hellwing:2017pmj,Du:2016aik} in galaxy as well as in  galaxy clusters. Also, chameleon effect has been introduced to explain Galactic rotation curves \cite{Burrage:2016yjm}. Here in our model,  interaction is such that in the small scale the scalar force takes over gravity and dominates the dynamics of the fermion-scalar system.  We show that when the scalar force is balanced by Fermi pressure, a static profile of the scalar field is obtained that makes the bound structure of fermions (DM) stable as it minimizes the action. From our static solution, we find that the scalar field starts from a high value near the center of the galaxy and reaches an asymptotic (near zero) value a few $~ \rm{kpc}$ away from it. As a result, dark matter is lighter towards the center and heavier away  from it and behaves like CDM at large distances. As dark matter becomes lighter due to smaller distance, one needs to take Fermi pressure into account and the stability of the system is obtained when Fermi pressure balances the attractive scalar force. By using the Thomas-Fermi approximation, we numerically solve for the scalar static profile $\phi(r)$, dark matter particle number density $n_{\psi}(r)$, and dark matter energy density $\rho_{\psi}(r)$ as a function of distance from the Galactic centre. For our reasonable choice of parameters, we show that dark matter density is naturally cored closer to the center of dSph galaxies.
    \par  The plan of the paper is as follows: in Sec. II we present the general setup for mass varying DM due to scalar interaction; in Sec. III, we provide a tentative particle physics scenario, describe the physics of the astrophysical system, derive the equations that need to be solved numerically. In Sec. IV, we explain numerical techniques and initial conditions along with the numerical results. In Sec. V, we discuss dSph galaxy observations in the context of our results and finally we conclude in Sec. VI.

\section{Mass varying DM:}
                
       \par Here we consider an interaction between dark matter particles and a chameleon scalar field ($\phi(r)$). We see that for our scenario, the mass of the chameleon scalar depends on the local matter density as well as the mass of the dark matter particles \cite{Casas:1991ky}.  Because of the interaction, dark matter particles  experiences a fifth force, an attractive force mediated by the scalar field. This attraction mediated by the chameleon scalar field may cause the dark matter particles to clump together and produce a compact stable dark matter halo when the attractive force is balanced by Fermi pressure. All of this physics  can be obtained  from a Lagrangian of a minimally coupled scalar with gravity along with the presence of a dark matter fermion whose mass depends on the scalar. \par
        For a real, classical scalar  field ($\phi$) minimally coupled to gravity, we can write down the action as \cite{Farrar:2003uw},
     \begin{equation}\label{S1}
     \mathcal{S}_{sca} = \int d^{4}x \sqrt{-g}[M^{2}R-\partial^{\mu}\phi\partial_{\mu}\phi - U(\phi)]
\end{equation}  Where  $M=(16\pi G)^{-1/2}$ and  $U(\phi)$ is the self-interacting scalar potential.
    On the other hand, action for spin-1/2 dark matter particle is \cite{Farrar:2003uw} given by
    \begin{equation}\label{S2}
    \mathcal{S}_{DMf} = \int d^{4}x\sqrt{-g}[ i \bar{\psi}\gamma^{\mu}\partial_{\mu}\psi-m_{\psi}(\phi) \bar{\psi}^{\mu}\psi_{\mu}]
    \end{equation}
     In the context of our scenario,  the De Broglie wavelength for the dark matter particle is much smaller than the characteristic length scale of variation of the scalar field. Hence, we can take the dark matter as classical gas of pointlike particles and the total action for our system boils down to  
    \begin{equation}\label{S4} 
    \begin{split}
    \mathcal{S}&=\mathcal{S}_{sca}+ \mathcal{S}_{DMf}\\
   & = \int d^{4}x \sqrt{-g}[M^{2}R-\partial^{\mu}\phi\partial_{\mu}\phi -V(\phi)]\\ &-\sum_{i}\int d\tau_{i}[m_{\psi}(\phi(x_{i}))]    
    \end{split}
    \end{equation}  

  To write down the Einstein equations one also needs the energy-momentum tensor and the metric. The energy-momentum tensor associated with the dark matter particles is\cite{Damour:1994zq}
    \begin{equation}\label{S5}
    T^{\mu\nu}=\frac{1}{\sqrt{-g}}\sum_{i}\int d\tau_{i}m_{\psi}(\phi(x_{i}))\frac{dx_{i}^{\mu}}{d\tau_{i}}\frac{dx_{i}^{\nu}}{d\tau_{i}}\delta^{(4)}(x-x_{i})
    \end{equation} 
      For the metric, we consider a static, spherically symmetric case ($\mathcal{M}=S_{2}\times U_{2}$, $S_{2}$ being a two-dimensional sphere and $U_{2}$ being a  two-dimensional metric with infinite manifold), so we can write the  space-time($\mathcal{M},g$) as
 \begin{align}\label{S8}
g = -A_{0}(r)dt^{2}+A_{1}(r)dr^{2}+r^{2}(d\theta^{2}+\sin^{2}\theta d\phi^{2})
\end{align} 
\par We consider the scalar field as a function of space only, not of time, as we are interested in the static configuration of the dark matter halo. Then, as done in \cite{Brouzakis:2005cj}, we first calculate  Einstein tensors for the above-mentioned metric and assume diagonal form of the energy-momentum tensor,   $T_{\mu}^{\mu}=3P-\rho$, treating dark matter as an isotropic fluid. Varying the action with respect to the scalar field in this static, spherically symmetric space-time and using the Einstein tensors we get
\cite{Brouzakis:2005cj} 
\begin{equation}\label{S9}
\begin{split}
&\phi^{\prime\prime}+\Big[\frac{1+A_{1}}{r}-\frac{A_{1}r}{2M^{2}}\Big(U(\phi)+\frac{1}{2}(\rho - P)\Big)\phi^{\prime}\\&=A_{1}\Big[\frac{dU}{d\phi}-\frac{d\ln m}{d\phi}T^{\mu}_{\mu}\Big]
\end{split}
\end{equation} where $\phi^{\prime}=\frac{d\phi}{dr}$.
For conservation of the energy-momentum tensor we get
\begin{equation}\label{S10}
\frac{dP}{dr}=\frac{d\ln m_{\psi}}{d\phi}\frac{d\phi}{dr} T^{\mu}_{\mu}-\frac{\rho +P}{2}\frac{d\ln A_{0}}{dr}
\end{equation}
These are two main equations that control the static solution of a scalar-fermion interacting system in the astrophysical context.

\section{Our Model}
\subsection{Weak limit of GR}

\par As we are interested in the dwarf galaxy dark matter halo, we work in the Newtonian limit of general relativity (GR),  $A_{0}\sim (1+2\Phi)$($|\Phi|\ll 1$, where $\Phi$ stands for Newtonian potential) and $ A_{1}\sim 1$.  With this approximation \eqref{S9} becomes
\begin{equation}\label{phi}
\begin{split}
&\phi^{\prime\prime}+\Big[\frac{2}{r}+\frac{\Phi^{\prime}}{(1+2\Phi)}\Big]\phi^{\prime}=\frac{dU}{d\phi}-\frac{d\ln m_{\psi}}{d\phi}T^{\mu}_{\mu}
\end{split}
\end{equation}
and from \eqref{S10} we get
\begin{equation}\label{p}
\frac{dP}{d\phi}=\frac{d\ln m}{d\phi} T^{\mu}_{\mu}-(\rho + P)\frac{\Phi^{\prime}}{(1+2\Phi)}
\end{equation}
 As we see in the Sec. V that the static solution allows the scalar field vacuum expectation value  to vary within \rm{kpc} from the dSph center and after this distance it drops to 0. So  within  this range, we find that the scalar force  between dark matter particles  is much more important than gravity. Hence, we can effectively take $A_{0}\Rightarrow 1$ and Eq. \eqref{phi} boils down to
\begin{equation}\label{phif}
\begin{split}
&\phi^{\prime\prime}+\frac{2}{r}\phi^{\prime}=\frac{dU}{d\phi}-\frac{d\ln m_{\psi}}{d\phi}T^{\mu}_{\mu}
\end{split}
\end{equation}
And the energy-momentum conservation equation simplifies to 
\begin{equation}\label{pf}
\frac{dP}{d\phi}=\frac{d\ln m_{\psi}}{d\phi} T^{\mu}_{\mu}
\end{equation}

In Sec. \ref{numerical}, we see that for  static structure of $\phi$ , the scalar field drops to 0 near $1 \rm{kpc}.$ As the dark matter is scalar field($\phi$) dependent, \eqref{pf} tells us that the attractive force is 0 for distance $ r \geq  1\,  \rm{kpc}$, thus reproducing GR in larger scales as stated before. \par 

\subsection{Particle physics ingredients}
\normalfont 
 The two main ingredients that we take as a toy low energy effective  model of the dark sector are  $ m_{\psi} \sim g / \phi $ and
 $U(\phi) \sim m^{2} \phi^{2}$, where $g$ has dimension $ [mass]^2$. We choose the two constants $ g, m $ appropriately for our numerical solution in next section  to get a cored DM halo in the sub-$\rm{kpc}$ scale. In the literature, people have taken different forms of $\phi$ dependent fermion mass and $m \simeq 1/\phi$ may also arise in dark sector or neutrino mass sector \cite{Brouzakis:2007aq, Amendola:2007yx,Fardon:2003eh}. 
 For a quadratic potential  $U(\phi) \simeq  m^2 \phi^2$, later we see  that the effective potential  $ U_{eff}=P-U $  has a minimum when one takes inverse power law coupling.  The presence of minima in the effective potential  is crucial to have a static structure; the reason behind this is discussed in the numerical section. That is  why we choose an inverse power law coupling as a toy model.   
 Now how this form of mass and potential arises from a fundamental  theory is beyond the scope of this work \cite{Halverson:2017deq}. Our goal here is to show that for such a scalar dependent dark matter mass and for a quadratic self-interacting potential, it is possible to get a static scalar profile that results in a stable dark matter halo in small dSph galaxies with cored density profile. We  later see that getting a static profile solution for this interacting fermion-scalar fluid needs many iterations to find an appropriate initial condition. So for another form of potential and scale dependent mass, whether such a solution can be obtained is beyond the scope of this work and has been kept for future research.

\subsection{$T^{\mu}_{\mu}$ for interacting dark matter-scalar  fluid}
Another final important piece we need  for our numerical solution is the functional form of the energy-momentum tensor of the dark matter cloud in the presence of attractive fifth force. 
Formation of compact fermionic objects, for example, soliton stars or fermionic Q-stars, in the presence of a scalar-mediated force has long been considered in different astrophysical contexts \cite{Lee:1986tr, Lynn:1989xb}, and more recently the formation of relativistic stars in chameleon
theories was considered \cite{Upadhye:2009kt}. But in our case, though inspired from the above work, the setup is quite different and  we expect dark matter particles to clump within a distance of the order of $1 \, \rm{kpc}$. For the scalar field $\phi$ value being high near the Galactic center, the mass of dark matter particles [$m_{DM} \sim 1/ \phi(r)$] will be very small there. Thus, the  dark matter Fermi particles have to obey the Pauli exclusion principle and they experience Fermi degeneracy pressure. This degeneracy pressure acts against the attractive fifth force. When these two forces balance each other, we obtain a static configuration for dark matter particles. 
\par Following \cite{Brouzakis:2005cj}, we use the Thomas-Fermi approximation to describe the distribution of the dark matter particles assuming weak interaction and no scattering. Then  dark matter particles at every point of space-time having local Fermi momentum $p(r)$  obey the distribution
\begin{equation}
  f(p)=\frac{1}{\exp\Big(\frac{\sqrt{p^{2}+m_{\psi}^{2}(\phi)}-\mu(r)}{T(r)}\Big)+1}
  \end{equation} The chemical potential $\mu$ and temperature $T$ is in local frame.  For simplicity we work in the zero temperature limit as in \cite{Brouzakis:2005cj}  and the expressions for density, pressure, and number density are given by 
  \begin{equation}\label{fermi}
  \begin{split}
  P(r)& =\frac{1}{12\pi^{3}}\int^{p_{F}}d^{3}p \frac{p^{2}}{\sqrt{p^{2}+m_{\psi}^{2}}}\\
  \rho(r)& = \frac{1}{4\pi^{3}}\int^{p_{F}}d^{3}p\sqrt{p^{2}+m_{\psi}^{2}}\\
  n & =  \frac{1}{4\pi^{3}}\int^{p_{F}}d^{3}p = \frac{p_{F}^{3}}{3\pi^{2}}
  \end{split}
  \end{equation}
Assuming the dark matter gas is in a sphere of radius $p_{F}$ in momentum space we get
 \begin{equation}\label{p1}
 \begin{split}
  P& = \frac{m_{\psi}^{4}}{4\pi^{2}}\Big[\frac{z^{3}_{F}\sqrt{1+z^{2}_{F}}}{3}-\frac{(z_{F}\sqrt{1+z^{2}_{F}}-\ln(z_{F}+\sqrt{1+z^{2}_{F}}))}{2}\Big]\\
  \rho&= \frac{m_{\psi}^{4}}{3\pi^{2}}z_{F}^{3}\sqrt{1+z^{2}_{F}} - P
  \end{split}
  \end{equation}where $z_{F}=\frac{p_{F}}{m_{\psi}}$.
  So, we get the energy-momentum tensor of the dark matter particles as,
  \begin{equation}\label{T1}
T^{\mu}_{  \mu} = -\rho +3P =\frac{m_{\psi}^{4}}{2\pi^{2}}[\ln(z_{F}+\sqrt{z^{2}_{F}+1})-z_{F}\sqrt{1+z^{2}_{F}}]
\end{equation}   
This functional form of $T^{\mu}_{  \mu}$ goes into the rhs of \eqref{pf}  and one can numerically find the $P(\phi)$ by solving  \eqref{pf} . Once we have a solution of $P(\phi)$, we are ready to solve \eqref{phif}.

\section{Numerical Solutions:}\label{numerical}
\begin{figure}[H]
\centering
\includegraphics[scale=0.16]{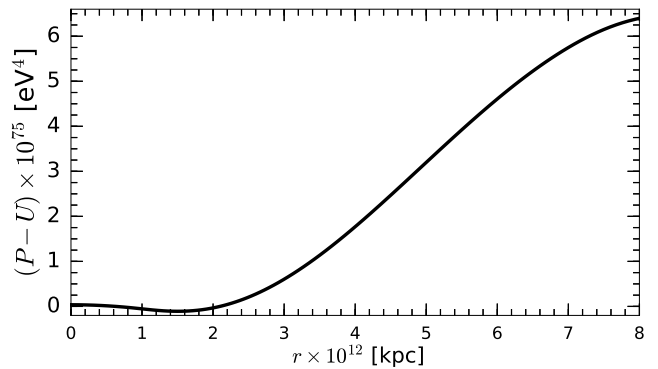}
\caption{ Numerically solved effective potential (P-U) as a function of $\phi$. Static solution is obtained  when $\phi$ is released from a high value and the field slowly stabilizes  around or at  the  minima of this effective potential.}\label{p-U}
\end{figure}

 Now we have all the necessary ingredients to solve for the static profile of the scalar field by numerically integrating  \eqref{phif} with two initial conditions.  One can rewrite the scalar equation  by combining \eqref{phif} and \eqref{pf} as 
\begin{equation}
\phi^{\prime\prime}+\frac{2}{r}\phi^{\prime}=-\frac{d(P-U)}{d\phi}
\end{equation}
If one can imagine that $r$ is replaced by time $t$ and $\phi$ is replaced by the position of the particle $x$, then the above equation represents a particle moving in Newtonian potential $(P-U)$ while the second term of the lhs represents the friction term. Now whether a static solution exits or not  depends on the shape of the effective potential. It is similar to the situation when a particle is released from some height of a potential and  the potential and friction allows the particle to be at rest at some other point. We plot our effective potential after getting $P(\phi)$ by solving  \eqref{pf}   in Fig.\ref{p-U} for suitable value of   $m_{\phi}^{2}=9.2 \times 10^{-53} \, \rm{eV}^{2}$ and $g=10^{-30.7}$.  Later we  see that this  choice of values in this range  gives us the desired static profile of scalar field that explains the dSph cored dark matter profile at observed distances. 

\par We realize  that quantum correction is an issue in such low mass scalar field situations. This is a challenge for all  low mass quintessence models also.  As here we are expecting  to solve the core-cusp issue in the dwarf spheroidal, the range of the fifth force has to be a few $\rm{kpc}$ or higher. This length scale somehow fixes  the $m_{\phi}$ to such a low value. Once m is fixed, we find that $g$ has to be also in a certain range ( which is also very tiny as chosen above) to balance the scalar force and quantum pressure. 
Now, how such a low mass scalar can be stable is beyond the scope of the paper. But we cite a few works where people deal with low mass scalars like the ULA etc. and some ideas or explanations are given about how a low mass scalar can indeed be stable against quantum correction. But the details of running for our case are kept for future work.

\begin{figure}[H]
\centering
\includegraphics[scale=0.15]{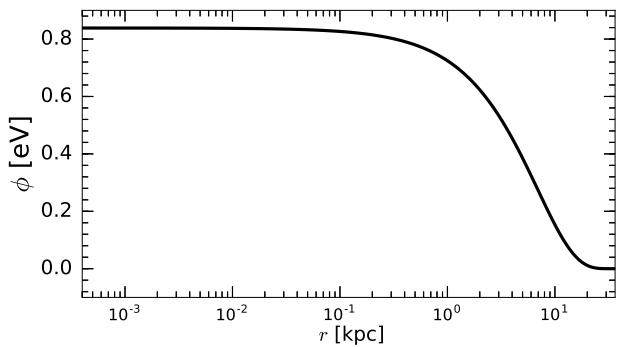}
\caption{ Static solution  of $\phi$. The plot shows a flat scalar field towards the centre and falling to 0 (very small value) at kpc order. }\label{phistatic}
\end{figure}

So getting a static solution is essentially an initial value problem in this situation. One has to find an appropriate fine-tuned  initial condition for which the scalar is static at asymptotic value. This initial value is chosen by iterating many times and finally achieving the static profile. First we start with choosing values of $m_{\phi}$ and $g$ that come from the Lagrangian as input parameters of the model. So  for our choice of  $m_{\phi}^{2}=9.2 \times 10^{-53} \, \rm{eV}^{2}$ and the coupling $g=10^{-30.7} \rm{eV}^2$, we look for a solution for $p_{F}$ by solving the force balance equation \eqref{pf} using \eqref{p1}  . Here we have given the appropriate initial condition $p_{F}[\phi_{initial}]=p_{F}^{0} \simeq 7.8  \times 10^{-35} \rm{eV}$ at $\phi_{initial}= 6.2 \times 10^{-13} \rm{eV}$. For this choice, the form of effective potential $(P-U)$ is shown in Fig.\ref{p-U}.  After getting  an expected form of the effective potential, one needs to solve \eqref{phif} for the scalar field static profile. It is a second order differential equation and hence we have to put two boundary conditions $\phi(r_{initial})=\phi_{0}$, and $\phi^{\prime}(r_{initial})=\phi^{\prime}_{0}$ in an interval  of $(r_{initial},r_{final})$. The idea for $\phi_{0}$ comes by studying the effective potential of Fig.\ref{p-U} carefully. As we have discussed in the comparison of our situation  with a particle moving in a Newtonian potential,  the scalar field  should be released from some $\phi_{0} $ for which the field moves towards  the minima to become static at the minima or around it due to frictional force. After many iterations we find that for  $\phi_{0}=8.1 \times 10^{-12} \, eV$ at $r_{initial}=3.47 \times 10^{27} \, \rm{eV}^{-1}$ and for $\phi_{0}^{\prime}=0$ one achieves a static profile for the scalar field $\phi(r)$, which we show in Fig.\ref{phistatic}.

\par
\begin{figure}[H]
\centering
\includegraphics[scale=0.15]{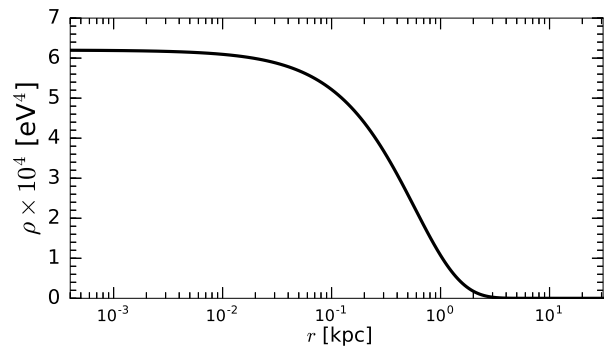}
\caption{ Mass density distribution  of dark matter particles inside the  halo. The flat density profile inside describes the cored profile of dark matter for the dSph galaxy at distance $r \leq \rm{kpc}$}\label{densityplot}
\end{figure}

\begin{figure}[H]
\centering
\includegraphics[scale=0.15]{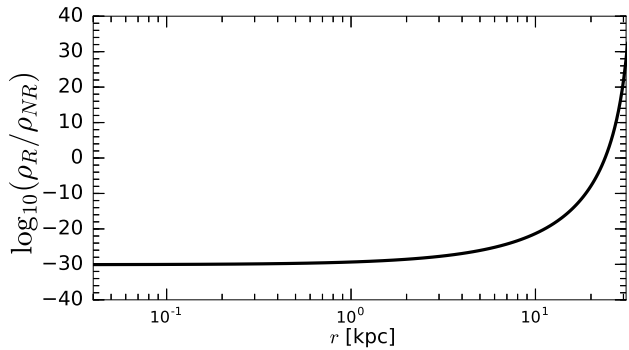}
\caption{ A comparison of the energy density between the relativistic and nonrelativistic case of dark matter particles.}\label{rhocompare}
\end{figure}

 As we can see, $\phi$ starts falling around $1 \, \rm{kpc}$ and falls close to 0 near $20 \, \rm{kpc}$. After obtaining the static profile the dark matter particle density can be calculated using \eqref{p1}. Using this formula we get density distribution for the dark matter halo as in Fig.\ref{densityplot}, which is the main result of our work. The dark matter energy density is found to be $6.2 \times 10^{-4} $ $eV^{4}$ near the core and then falls near 0 between  $0.1-5$ $\rm{kpc}$. We also  find that the dark matter mass increases outwards and reaches a maximum value of the order of $ \, \rm{MeV} $ to $\rm{GeV}$. 
We also obtain number density of dark matter particles  using $n=\frac{p_{F}^{3}}{3\pi^{2}}$ as given in \eqref{fermi}.
We find that the dark matter number density also falls to 0 at  $ \sim \rm{kpc}$ and its value inside the halo is $ \simeq 2.09 \times 10^{-3} \rm{eV}^{3}$.
As expected,  the dark matter particles are not nonrelativistic near the Galactic center as the mass is lighter  and one has to use the general formula for dark matter energy density as given in \eqref{p1}.  We find that at  larger distance the dark matter energy density slowly maps to nonrelativistic case like the CDM. This can be seen from to the ratio plot Fig.\ref{rhocompare} between relativistic and nonrelativistic case which ensures that we recover the CDM paradigm at larger distance, thus  recovering the success of large scale structure formation.
  We also clearly see that dark matter density does not rise at $1/r$ towards the center rather it flattens, thus giving a cored density profile naturally.
For our appropriate choice of parameters, it is  also possible to get the dark matter density
 $\sim 10^{-4} \, eV^{4}$ 
 within the dark matter halo, which matches recent dSph galaxy dark matter observations.

\section{Cored Structure of dSph galaxies:}

\subsection{Comparison of our numerical results and dSph observations }
The dSph satellite galaxies are the smallest and faintest galaxies observed till now, are known to be dark matter dominated at all radii, and they have the largest dynamical mass-to-light ratios($[M/L_{V}]/[M/L_{V}]_{\odot}\gsim 10^{1-2}$)\cite{Walker(2011), Gonzales-Morales:2016mkl}. As baryonic mass content is much less in these galaxies, uncertainties in determining the baryonic mass profile have very little effect on the determination of the dark matter mass profile \cite{Walker(2011)}. According to observation, for dSph galaxies the dark matter mass  
distribution has a core structure (that is, dark matter energy density $\rho_{DM}\sim r^{0}$). But N-body simulations in the collisionless $\Lambda$CDM paradigm produce the cusp structure of dark matter mass distribution (dark matter energy density $\rho_{DM}\sim r^{-1}$)\cite{deblok(2010)}.

In our work the balance between the proposed attractive fifth force and the Fermi degeneracy pressure between the dark matter particles has produced a static profile of the scalar field and hence 
a core profile of dark matter energy density. According to \cite{Hui:2016ltb}, for 36 local group dSph galaxies maximum, mean, and median value of dark matter energy density is respectively $1.5\times10^{-3}$, $1.5\times 10^{-4}$ and $3\times10^{-5}$ in $\rm{eV}^{4}$. The core profile of dark matter energy density we have produced has a value of $7.988 \times 10^{-4} $ $\rm{eV}^{4}$ within $\sim  0.1$ $\rm{kpc}$, which agrees with the present scenario of observations and simulations on the core structure of dSph mass distribution.
\par  We hence show that in the presence of a chameleon scalar field, it is possible to produce a core dark matter mass distribution profile like dSph galaxies with the assumption that the attractive fifth force dominates over gravitation within the scale of dark matter mass distribution for the dSph galaxies.  Along with light dark matter particles at small scale (which produce the core profile), outside the length scale of dSph galaxies we get heavy dark matter particles($\sim MeV$) reproducing the $\Lambda$CDM paradigm. There was recent work  done in \cite{Randall:2016bqw}, where the pressure of the quasidegenerate fermi dark matter gas is balanced by the self-gravitation of the dark matter particles. They have produced a dSph core profile of size $\ge 130$ $\rm{pc}$ for constant dark matter masses in the range $70- 400$ $\rm{eV}$.

\begin{figure}[H]
\centering
\includegraphics[scale=0.35]{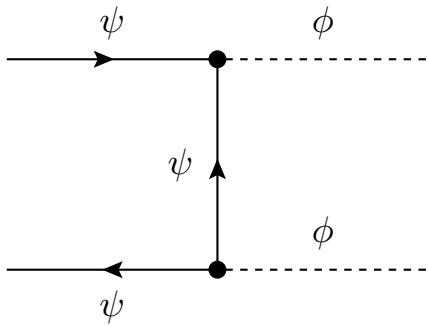}
\caption{ Feynman diagram of dark matter annihilation into scalar radiation.}\label{fd}
\end{figure}

\subsection{Stability of dark matter halo and $\phi $ radiation }

As the scalar couples to the fermion, there is a possibility of dark matter decay into scalar radiation. Though dark matter becomes very light towards the Galactic  center, its mass increases  considerably and rate of  $\phi $ radiation through  \, $  \psi + \psi \rightarrow \phi + \phi $\, \, process. Here we estimate the rate and show that the dark matter halo is stable  compared to the age of the Universe. 
 We show that the nuggets  are stable enough to be dark matter by calculating  its decay rate into $\phi$.
The coupling between scalar and dark matter enters through the $\phi $ dependent dark matter mass $ ( g / \phi ) \, \psi \, \psi$.   
Inside the halo, for small fluctuation $\delta \phi$  this  can be written as 
$  \frac{dm_{\psi}}{d\phi} \mid_{_{\phi_{static}}} \delta \phi \, \psi \, \psi$
 and we  find that coupling constant  $\kappa_{in}$ inside the halo is given by 
 
 \begin{equation}
 \kappa_{in} \simeq   \frac{g}{\phi(r)^2}
 \end{equation}
 
 where $\phi(r)$ has to be  taken from our numerical static solution for the scalar. We find that for our numerical solution the coupling is extremely tiny, $\kappa \simeq  10^{-28} - 10^{-30} $, inside the halo. 
The nugget lifetime  can be estimated for a given $\kappa$. If n is the number density of the fermion inside the nugget, the decay rate is given by
 $\frac{dn}{dt} \simeq n^{2} \frac{ \kappa^{4} }{32 \pi E^{2}_{CM}}$. 
Integrating this, we find an estimate for the half life of the nugget, \,  
$ \Delta t_{1/2} \simeq \frac{1}{n (\frac{ \kappa^{4} }{32 \pi E^{2}_{CM}})}$.
 Substituting the values from our numerical solution, we find that the half life  of the nugget is roughly  $ \Delta t_{1/2} \sim 10^{142-146} $Sec. This is way greater than the age of the universe ($ t_U \sim 10^{17}$ Sec). 
 
\subsection{Strength of fifth force compared to gravity }
Once we have the coupling $\kappa (r)$,  it is easy to estimate the strength of the fifth force between dark matter particles as a function of distance. Following \cite{Bovy:2008gh}, the strength ratio $\beta$ between fifth force and gravity is given by  $\beta \simeq \frac{\kappa(r)^2 / (4 \times \pi \times m_{\psi}^2(r))}{G_N}$. We find that the value of $\beta \gg 1 $ for distance $r \leq 0.1$ $\rm{kpc}$. So, the fifth force is much stronger than gravity inside the halo,  which validates our approximation for neglecting gravitational potential ( which is equivalent of taking $A_0 \rightarrow 1 $ ) in Eq. (10).  Also, it is instructive to note that the limit on the strength of dark matter fifth force derived in \cite{Kesden:2006vz} from tidal disruption of the satellite galaxy does not apply for our case, as the range of the force is much larger there. In our case the coupling constant $\kappa \propto \frac{dm_{\psi}}{d\phi}$, so the strength $\kappa \rightarrow 0$ as $m_{\psi}$ becomes constant when the scalar achieves an asymptotic constant value at a distance $\leq$ few $\, \rm{kpc}$.
\par Till now , to be safe, we have assumed that the scalar only couples to dark matter.  If it couples to the baryon also, then the situation is complicated and  whether a static structure of the baryon cloud could be formed with $\rm{kpc}$ range force in  the Milky Way galactic disk, needs detailed study as the baryon has other interactions. This has been kept for future work and to avoid a local test of gravity constraints on fifth force, we have assumed that the dark matter only couples to the scalar field in the dark sector. But it is instructive to note that, in a high density environment, like in the Solar System, if there is  any static structure of the baryon at all due to this fifth force, we have checked from our numerical code that the scalar field value would be much higher in the high matter density environment. We have also checked that the fifth force strength goes down rapidly for the high scalar field value, so we are moving in the right direction to evade solar system constraints. But as we do not know whether baryonic static structure can be formed (as it has other interactions and in the Milky Way disk the problem turns out to be much more complicated), we assume that the baryon does not experience this scalar-mediated  fifth force for this  work. 

\section{discussion}
Though CDM cosmology is amazingly successful in its prediction in large scale observations like CMB, BAO, and LSS, but small scale galactic observations are  incompatible for many CDM predictions. The core vs cusp problem in dwarf galaxies is one such issue that remains one of the strongest challenges to the CDM paradigm. These small dwarf galaxies are dark matter rich, so even the baryonic feedback (which rescues other small scale CDM N-body issues) would not do a great job due to lack of baryons in dwarf galaxies. Recently, a solitonic cored profile of ultralight scalar dark matter \cite{Marsh:2015wka, Hui:2016ltb} was proposed as a physical explanation of the cored DM profile. But within CDM, there exists no solid physical explanation for a cored profile of dark matter towards the center of these dSph galaxies. Here, for the first time, we provide a possible physical explanation for CDM to form a cored density profile through small scale modification of gravity in the presence of scalar fifth force in the dark matter sector. Because of variation of the scalar field profile towards the center of dSph, dark matter mass becomes lighter and Fermi pressure starts to balance the fifth force, giving a static configuration of the dark matter cloud. As the scalar field value flattens towards the center, the dark matter density, which is a function of scalar field profile, tends to flatten towards the center, naturally giving a cored profile.  
Also, it is instructive to note that the scalar field value asymptotically drops to 0 near $ \simeq $few $\rm{kpc}$. As the dark matter mass is scalar field($\phi$) dependent and inversely proportional to $\phi(r)$, naturally we see that dark matter behaves like CDM far away from the center of the galaxy. So while putting our setup in N-body simulation, at larger distance once can safely use the normal N-body recipe for simulations while at distances less than $\rm{kpc}$ one needs to take into account this new scalar force and Fermi pressure. That work is beyond the scope of the present work and is kept for future research. 
\section*{ACKNOWLEDGEMENTS}
We thank Kris Sigurdson and Neal Weiner for initial  helpful discussions at the beginning of the project.  S. D. thanks CCPP, NYU for giving an opportunity to visit where considerable progress was made. 

\end{document}